\documentclass{PoS}

\usepackage[numbers]{natbib}

\title{Recent Progress on Ascertaining the Core Collapse Supernova Explosion Mechanism}

\ShortTitle{Core Collapse Supernova Mechanism}
        
\author{
	\speaker{Anthony Mezzacappa}$^{ab}$,
	Stephen W. Bruenn$^{c}$,
	Eric J. Lentz$^{abd}$,
	W. Raphael Hix$^{ad}$,
	
	J. Austin Harris$^{a}$,
	O. E. Bronson Messer$^{ade}$,
	Eirik Endeve$^{abf}$,
	Merek A. Chertkow$^{a}$,
	
	John M. Blondin$^{g}$,
	Pedro Marronetti$^{h}$, and
	Konstantin N. Yakunin$^{abd}$
		\thanks
		{	
			This research was supported by the U.S. Department of Energy Offices of Nuclear Physics; the NASA Astrophysics Theory and Fundamental Physics Program (grants NNH08AH71I and NNH11AQ72I); and the National Science Foundation PetaApps Program (grants  OCI-0749242, OCI-0749204, and OCI-0749248).
			PM is supported by the National Science Foundation through its employee IR/D program. The opinions and conclusions expressed herein are those of the authors and do not represent the National Science Foundation.
			This research was also supported by the NSF  through TeraGrid resources provided by the National Institute for Computational Sciences under grant number TG-MCA08X010; resources of the National Energy Research Scientific Computing Center, supported by the U.S. DoE Office of Science under Contract No. DE-AC02-05CH11231; and an award of computer time from the Innovative and Novel Computational Impact on Theory and Experiment (INCITE) program at the Oak Ridge Leadership Computing Facility, supported by the  U.S. DOE Office of Science under Contract No. DE-AC05-00OR22725.
		}\\
    \llap{$^a$}
    	Department of Physics and Astronomy, University of Tennessee,
	    Knoxville, TN 37996, USA\\
	\llap{$^b$}
		Joint Institute for Computational Sciences, ORNL,
		Oak Ridge, TN, 37831 USA\\
	\llap{$^c$}
		Department of Physics, Florida Atlantic University,
		Boca Raton, FL 33431, USA\\	    
	\llap{$^d$}
	    Physics Division, Oak Ridge National Laboratory,
	    Oak Ridge, TN 37831, USA\\
	\llap{$^e$}
		National Center for Computational Sciences, ORNL,
		Oak Ridge, TN 37831, USA\\
	\llap{$^f$}
		Computer Science and Mathematics Division, ORNL,
		Oak Ridge, TN 37831, USA\\	
	\llap{$^g$}
		Department of Physics, North Carolina State University,
		Raleigh, NC 27695, USA\\
	\llap{$^h$}
		Physics Division, National Science Foundation,
		Arlington, VA 22230, USA\\
    E-mail: \email{mezz@utk.edu}
}

\abstract{
We have been working within the fundamental paradigm that core collapse supernovae (CCSNe) may be neutrino driven, since the first suggestion of this by Colgate and White nearly five decades ago. Computational models have become increasingly sophisticated, first in one spatial dimension assuming spherical symmetry, then in two spatial dimensions assuming axisymmetry, and now in three spatial dimensions with no imposed symmetries. The increase in the number of spatial dimensions has been accompanied by an increase in the physics included in the models, and an increase in the sophistication with which this physics has been modeled. Computation has played an essential role in the development of CCSN theory, not simply for the obvious reason that such multidimensional, multi-physics, nonlinear events cannot possibly be fully captured analytically, but for its role in discovery. In particular, the discovery of the standing accretion shock instability (SASI) through computation about a decade ago has impacted all simulations performed since then. Today, we appear to be at a threshold, where neutrinos, neutrino-driven convection, and the SASI, working together over time scales significantly longer than had been anticipated in the past, are able to generate explosions, and in some cases, robust explosions, in a number of axisymmetric models. But how will this play out in three dimensions? Early results from the first three-dimensional (3D), multi-physics simulation of the "Oak Ridge" group are promising. I will discuss the essential components of today's models and the requirements of realistic CCSN modeling, present results from our one-, two-, and three-dimensional models, place our models in context with respect to other efforts around the world, and discuss short- and long-term next steps.
}

\FullConference{The 32nd International Symposium on Lattice Field Theory,\\
		23-28 June, 2014\\
		Columbia University, New York, NY}

\newcommand{\chimera}{\textsc{Chimera}}
\newcommand{\AgileBOLTZTRAN}{\textsc{Agile-BOLTZTRAN}}
\newcommand{\vertex}{\textsc{PROMETHEUS-VERTEX}}
\newcommand{\coconutvertex}{\textsc{COCONUT-VERTEX}}
\newcommand{\msun}{\ensuremath{M_{\odot}}}
\newcommand{\nue}{\ensuremath{\nu_{e}}}
\newcommand{\nuebar}{\ensuremath{\bar \nu_e}}
\newcommand{\numt}{\ensuremath{\nu_{\mu\tau}}}
\newcommand{\numtbar}{\ensuremath{\bar \nu_{\mu\tau}}}
\newcommand{\numu}{\ensuremath{\nu_{\mu}}}
\newcommand{\nutau}{\ensuremath{\nu_{\tau}}}
\newcommand{\numubar}{\ensuremath{\bar \nu_{\mu}}}
\newcommand{\nutaubar}{\ensuremath{\bar \nu_{\tau}}}
\newcommand{\mev}{\mbox{MeV}}
\newcommand{\gcc}{\ensuremath{{\mbox{g~cm}}^{-3}}}
\newcommand{\isotope}[2]{$^{#2}$#1}
\newcommand{\degree}{\ensuremath{^\circ}}
\newcommand{\Ediag}{\ensuremath{E^{+}}}
\newcommand{\Ediagov}{\ensuremath{E^{+}_{\rm ov}}}
\newcommand{\Ediagovrec}{\ensuremath{E^{+}_{\rm ov, rec}}}

\begin{document}

\section{The Physics and Status of Core Collapse Supernova Simulations}\label{sec:foundation}

Core collapse supernovae (CCSNe) are initiated by the collapse of the iron cores of massive stars at the ends of their lives. The collapse proceeds to ultrahigh densities, in excess of the densities of nucleons in the nucleus of an atom (``super-nuclear'' densities). The inner core becomes incompressible under these extremes, bounces, and, acting like a piston, launches a shock wave into the outer stellar core. This shock wave will ultimately propagate through the stellar layers beyond the core and disrupt the star in a CCSN explosion. However, the shock stalls in the outer core, losing energy as it plows through it, and exactly how the shock wave is revived remains an open question, although, important progress is being made, particularly with two-dimensional (2D) models. The means to this revival is the central question in CCSN theory. (For a more complete review, the reader is referred to \cite{Mezz05}, \cite{Jank12}, and \cite{KoTaSu12}.) 

After core bounce, $\sim10^{53}$~ergs of energy in neutrinos and antineutrinos of all three flavors  are released from the newly formed proto-neutron star (PNS) at the center of the explosion. 
The typical observationally estimated CCSN explosion energy  is $\sim 10^{51}$~ergs ($\equiv1$~Bethe), with estimates for individual supernovae ranging from 0.3--5~Bethe \cite{Hamu03,NoToUm06,Smar09}.
Past simulations \cite{Wils85,BeWi85} demonstrated that energy in the form of neutrinos emerging from the PNS can be deposited behind the shock and may revive it. 
This neutrino reheating is central to CCSN models today. 
However, while a prodigious amount of neutrino energy emerges from the PNS, the neutrinos are weakly coupled to the material below the shock. 
The neutrino heating is very sensitive to the distribution of neutrinos in energy (or frequency) and direction of propagation, at any given spatial point behind the shock 
\cite{BuGo93,JaMu96,MeMeBr98,MeCaBr98b,MeLiMe01,Jank01}.
Realistic CCSN simulations require a neutrino transport method that can reproduce the angular and energy distributions of the neutrinos in the critical heating region.

Normal iron core stars do not explode when modeled in spherical symmetry  \cite[cf.,][]{LiMeTh01a,RaJa02,ThBuPi03}, thus multidimensional effects are required.  
Fluid instabilities ({\it e.g.}, convection) in the PNS may boost the luminosity of this central neutrino source and consequent neutrino heating \cite{SmWiBa81,WiMa93,MiPoUr02,BrRaMe04,BuRaJa06}.  
Neutrino-driven convection between the PNS and the shock fundamentally alters the nature of energy flow and shock revival \cite{HeBeHi94,BuHaFr95,JaMu96,FrWa04,BuRaJa06,BrDiMe06} relative to the spherically symmetric case, allowing simultaneous down-flows that fuel the neutrino luminosities and neutrino-heated up-flows that bring energy to the shock. 
The standing accretion shock instability (SASI), a computationally discovered instability of the shock wave itself \cite{BlMeDe03}, dramatically alters the shock and explosion dynamics 
\cite{BlMeDe03,JaBuKi05,BuLiDe06,OhKoYa06,HaMuWo13}. Recent axisymmetric (2D) models \cite{MuJaHe12,BrMeHi13} demonstrate that neutrino heating in conjunction with neutrino-driven convection and the SASI are able to generate explosions, although the quantitative predictions --- in particular, the explosion energies --- differ between these two groups. However, it is important to note that our predictions are consistent with observations \cite{BrLeHi14} across a range of observables: explosion energy, $^{56}$Ni mass, neutron star mass, and neutron star kicks.
Despite these differences, these advances suggest that the SASI  may be the ``missing link'' that will enable the Wilson delayed-shock, neutrino-heating mechanism to operate successfully in multiple spatial dimensions, especially for more massive progenitors. 

There are many other inputs to the physics of the core collapse supernova (CCSN) mechanism that must also be included in simulations. The strength of these effects have been tested in many one-dimensional (1D) simulations and some multidimensional simulations.
The PNS in a developing CCSN is sufficiently compact to require the inclusion of general relativistic effects to gravity and neutrino propagation \cite{BaCoKa85,LiMeTh01a,LiMeTh01b,BrDeMe01,MaDiJa06,OtDiMa07,MuJaDi10,LeMeMe12a,MuJaMa12}.
Getting the correct radiative coupling requires inclusion of all neutrino--matter interactions (opacities) that affect the neutrino transport, heating, and cooling. Several recent studies have considered the effects of neutrino opacities, including inelastic scattering of neutrinos on electrons, nucleons, and nuclei, detailed nuclear electron capture, and nuclear medium effects on the neutrino interactions \cite{HiMeBr03,BuJaKe03,KeRaJa03,ThBuPi03,MaJaBu05,MaLiFr06,LaMaMu08,JuLaHi10,RoReSh12,LeMeMe12b}.
A nuclear equation of state for both nuclear matter in the PNS and the nuclei and nucleons in the surrounding matter is required. Several equations of state have been proposed \cite{BeBrAp79,ElHi80,Coop85,LaSw91,WiMa93,ShToOy98b,HeSc10,ShHoTe11,StHeFi13} and their impact in CCSNe has been examined \cite{SwLaMy94,RaBuJa02,SuYaSu05,MaJaMu09,LeHiBa10,Couc13a}.
Finally, the nuclear composition must be evolved in the outer regions where nuclear statistical equilibrium (NSE) does not apply.

The centrifugal effects of stellar core rotation, especially for rapid rotation, can also change supernova dynamics qualitatively and quantitatively \cite{FrWa04,BuRaJa06}. 
An additional level of complexity is added by models with dynamically important magnetic fields, amplified by rapid rotation and the magnetorotational instability, that may  play a significant role in driving, and perhaps collimating, some CCSNe \cite{Symb84,AkWhMe03,BuDeLi07} and \emph{collapsars} (jets generated by accretion disks about newborn black holes producing combined CCSNe/$\gamma$-ray bursts). 
Recent  observations of shock breakout \cite{ScJuWo08} disfavor a strongly collimated jet as the driver for explosions for ordinary supernovae \cite{CoWhMi09} --- i.e., cases where rotation likely does not play a major role. 
Magnetic fields are expected to become important in the context of rapidly rotating progenitors, where significant rotational energy can be tapped to develop strong and organized magnetic fields (e.g., see \cite{BuDeLi07}). State-of-the-art stellar evolution models for massive stars \cite{wohe07} do not predict large core rotation rates. For non-rapidly rotating progenitors, magnetic fields are expected to serve more of a supporting role, for neutrino shock reheating (e.g., see \cite{ObJaAl14}).
 
While the list of major macroscopic components in any CCSN clearly indicates this is a 3D phenomenon, 3D studies have been relatively rare and, until recently, generally have skimped, largely for practical reasons, on key physics to which prior studies (noted above) have indicated careful attention must be paid.  
3D simulations have examined aspects of the CCSN problem using a progression of approximations.
3D, hydrodynamics-only simulations of the SASI, which isolate the accretion flow from the feedbacks of neutrino heating and convection, have identified the spiral ($m=1$) mode, with self-generated counter-rotating flows that can spin the PNS to match the $\sim$50~ms periods of young pulsars \cite{Blon05a,Blon05b,BlMe07} and examined the generation of magnetic fields \cite{EnCaBu10} and turbulence \cite{EnCaBu12} by the SASI.
Another often-used formulation for approximate 3D simulations is the neutrino ``lightbulb'' approximation, where a proscribed neutrino luminosity determines the heating rate, with the neutrino heating and cooling parameterized independently. 
Neutrino lightbulb simulations have been used successfully to study the development of NS kicks \cite{NoBrBu12,WoJaMu12,WoJaMu13}, mixing in the ejecta \cite{HaJaMu10}, and, in 2D simulations, the growth of the SASI with neutrino feedbacks \cite{ScJaFo08}. Lightbulb simulations have also been used to examine the role of dimensionality (1D-2D-3D) in CCSNe \cite{MuBu08,NoBuAl10,HaMaMu12,Couc13b}.
A more sophisticated approximate neutrino transport method is the ``leakage'' scheme. Leakage schemes use the local neutrino emission rate and the opaqueness of the overlying material to estimate the cooling rate and from that the neutrino luminosity and heating rate. 
Leakage models have been used by Ott et al. \cite{OtAbMo13}, including the full 3D effects of GR.
Fryer and Warren \cite{FrWa02,FrWa04} employed a \emph{gray} neutrino transport scheme in three dimensions. In such schemes, one evolves the spatial neutrino energy and momentum densities with a 
parameterization of the neutrino spectra. As a neutrino angle- and energy-integrated scheme, the dimensionality of the models is greatly reduced, which is ideal for performing a larger number of exploratory studies.
These 3D studies, and other recent studies \cite[cf.][]{TaKoSu12,BuDoMu12,HaMuWo13,CoOc13}, confirm the conclusion that CCSN simulations must ultimately be performed in three spatial dimensions.  

The modeling of CCSNe in three dimensions took an important step forward recently. The Max Planck (MPA) group launched the first 3D CCSN simulation with multifrequency neutrino transport with relativistic corrections and state-of-the-art neutrino opacities, and general relativistic gravity. Results from the first 400 ms after stellar core bounce were reported in \cite{HaMuWo13} for a 27 \msun\ progenitor. At present, the ``Oak Ridge'' group is performing a comparable simulation beginning with the 15~\msun\ progenitor used in our 2D studies. We have evolved approximately the first half second after bounce (for further discussion, see Section~\ref{sec:current3D}). 

\section{Lessons from Spherical Symmetry}

\begin{figure}
\includegraphics[width=3.00in]{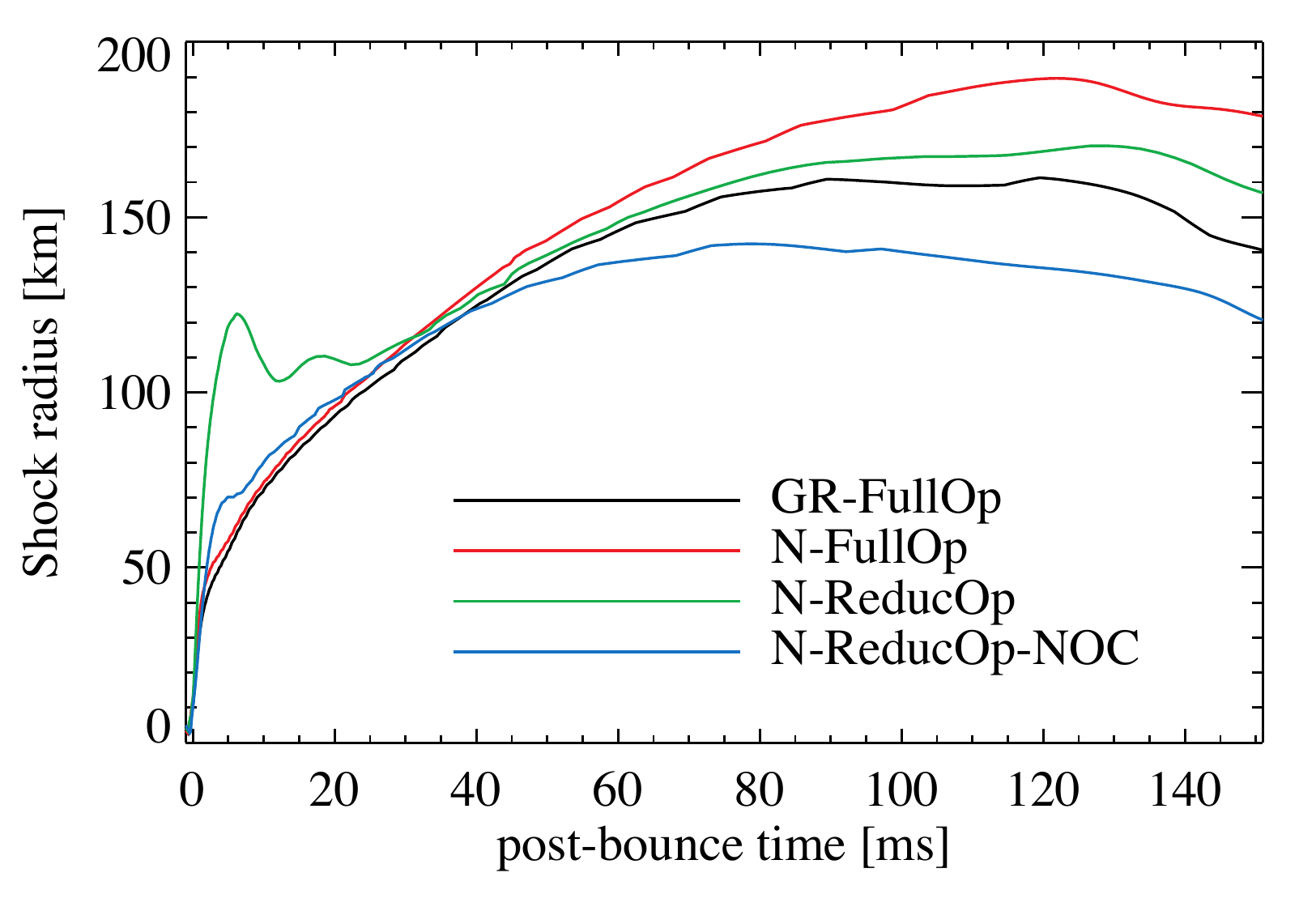}
\caption{Shock trajectories in km, versus time after bounce, for models with decreasing physics \cite{LeMeMe12}.}
\label{fig:shockvphysics}
\end{figure}

\begin{figure}
\includegraphics[width=3.00in]{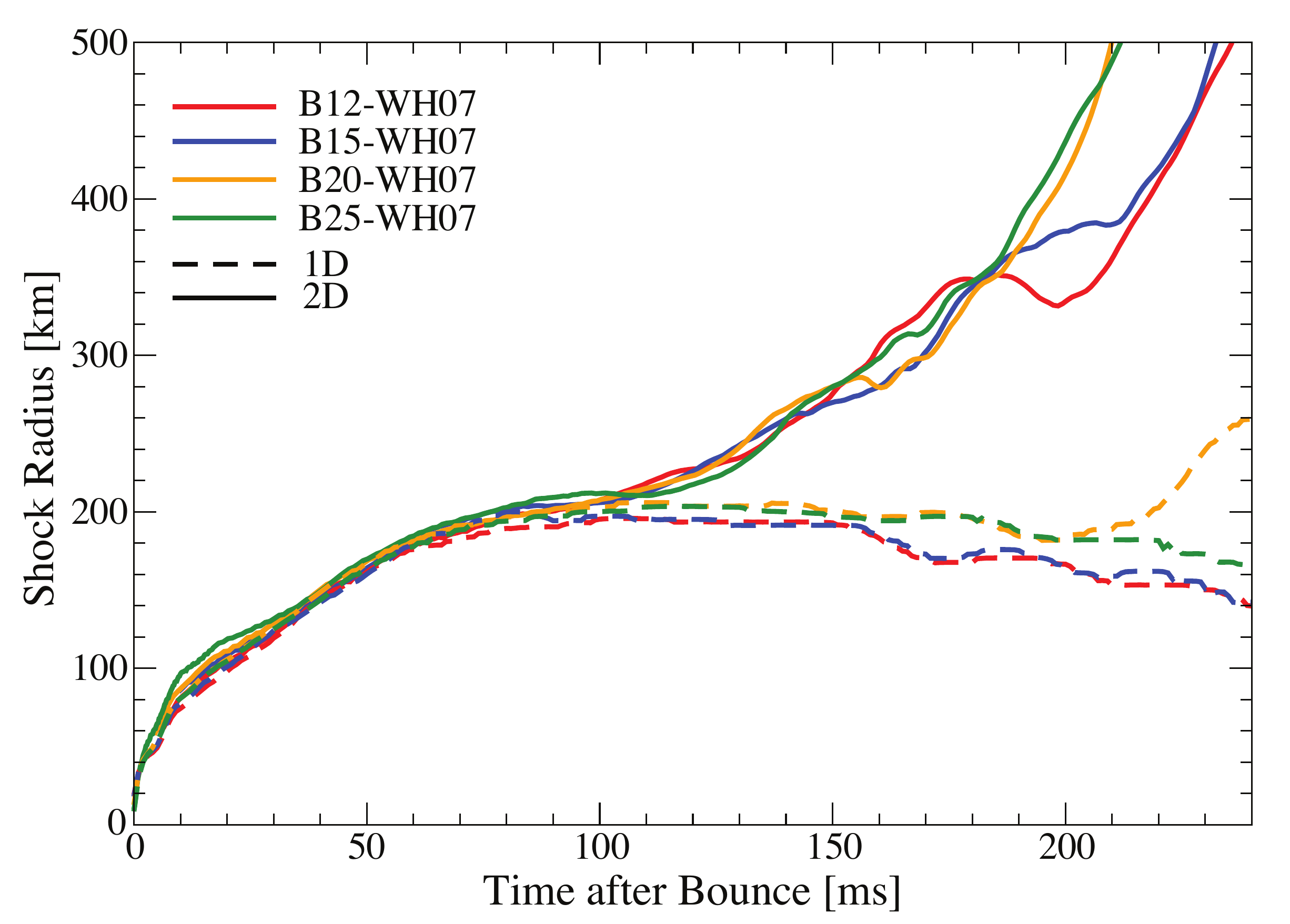}
\caption{Shock trajectories in km, versus time after bounce, for our 1D and 2D models \cite{BrMeHi13}. The 1D and 2D evolution begins to diverge between 100 and 125 ms after bounce.}
\label{fig:label1Dv2D}
\end{figure}

Recent studies carried out in the context of general relativistic, spherically symmetric CCSN models with Boltzmann neutrino transport demonstrate that (i) a general relativistic treatment of gravity, (ii) special and general relativistic corrections to the neutrino transport, such as the gravitational redshift of neutrinos, and (iii) the use of a complete set of weak interactions and a realistic treatment of those interactions are indispensable \cite{LeMeMe12a}. As shown in Figure \ref{fig:shockvphysics}, the impact of moving to a Newtonian description of gravity from a fully general relativistic treatment has a significant impact on the shock trajectory. The Newtonian simulation neglects general relativity in the description of gravity {\it per se}, as well as general relativistic transport effects such as gravitational redshift. Thus, the switch from a general relativistic description to a Newtonian description impacts more than just the treatment of gravity. In turn, if we continue to simplify the model, this time reducing the set of weak interactions included and the realism with which these weak interactions are included, we see a further significant change in the shock trajectory, with fundamentally different behavior early on after bounce. In this instance, we have neglected the impact of nucleon correlations in the computation of electron capture on nuclei (see \cite{HiMeMe03}), energy exchange in the scattering of neutrinos on electrons, corrections due to degeneracy and nucleon recoil in the scattering of neutrinos on nucleons, and nucleon--nucleon bremsstrahlung. Finally, if we continue to simplify the neutrino transport by neglecting special relativistic corrections to the transport, such as the Doppler shift, we obtain yet another significant change. The spread in the shock radii at $t>$120 ms after bounce is approximately 60 km. Its relative fraction of the average of the shock radii across the four cases at $t>$ 120 ms is $>$33\%. Moreover, the largest variation in the shock radii in our 2D models is obtained at $\sim$ 120 ms after bounce, which is around the time when the shock radii in our one- and two-dimensional models begin to diverge (see Figure \ref{fig:label1Dv2D}). In all four of our 2D models, the postbounce evolution is quasi-spherical until $\sim$110 ms after bounce. Thus, the use of the \AgileBOLTZTRAN\ code, which solves the general relativistic Boltzmann equation with a complete set of neutrino weak interactions for the neutrino transport in the context of spherically symmetric models, to determine the physics requirements of more realistic two- and three-dimensional modeling is possible. Indeed, the conclusions of our studies are corroborated by similar studies carried out in the context of 2D multi-physics models \cite{MuJaMa12}. Taken together, these studies establish the {\it necessary} physics that must be included in CCSN models in the future. Whether or not the current treatments of this physics in the context of two- and three-dimensional models is {\it sufficient}, as we will discuss, remains to be determined.

\section{Our Code}

\begin{figure}
\includegraphics[width=3.00in]{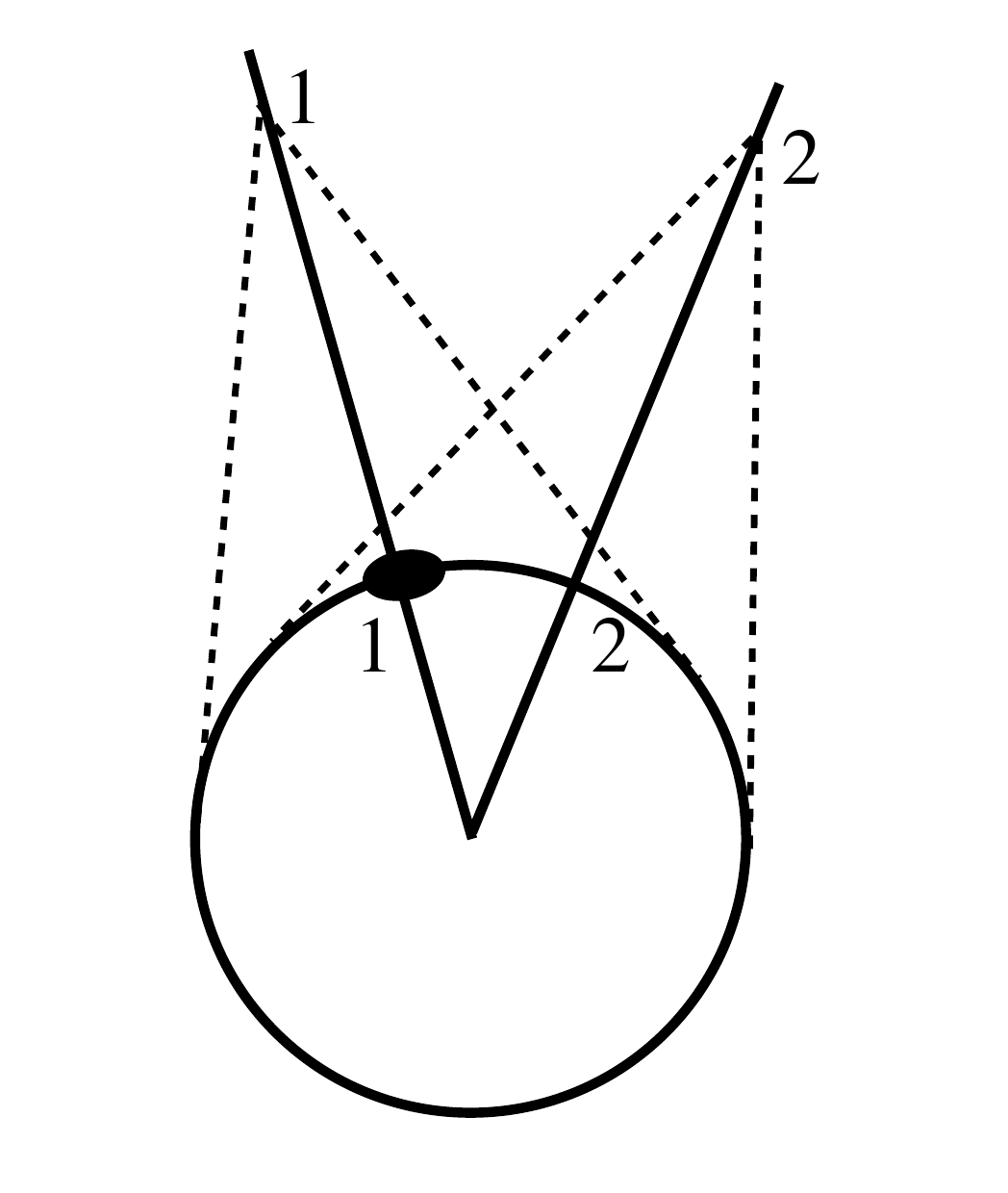}
\caption{A depiction of the ``ray-by-ray'' (RbR) approach. Each ray corresponds to a separate spherically symmetric problem. In the limit of spherical symmetry, the RbR approach is exact. Each ray solve gives what would be obtained in a spherically symmetric solve for conditions at the base of the ray, on the proto-neutron star surface. For a {\it persistent} hot spot, such as the one depicted here at the base of ray 1, the RbR approximation would overestimate the angular variations in the neutrino heating at the points 1 and 2 above the surface. In spherical symmetry, the condition at the base of each ray is assumed to be the same over the entire portion of the surface subtended by the backward causal cone for that ray. Thus, for ray 1, the entire subtended surface would be considered hotter than it is, whereas for ray 2 the contribution from the hot spot at the base of ray 1 to the heating at point 2 above the surface would be ignored.
\label{fig:rbr}
}
\end{figure}

\chimera\ is a parallel, multi-physics code built specifically for multidimensional simulation of CCSNe.
It is the chimeric combination of separate codes for hydrodynamics and gravity; neutrino transport and opacities; and a nuclear EoS and reaction network, coupled by a layer that oversees data management, parallelism, I/O, and control.

The hydrodynamics are modeled using a dimensionally-split, Lagrangian-Remap (PPMLR) scheme  \cite{CoWo84} as implemented in VH1 \cite{HaBlLi12}.
Self-gravity is computed by multipole expansion \cite{MuSt95}.
We include the most important effects of GR by replacing  the Newtonian monopole term  with a GR monopole computed from the TOV equations \cite[][Case~A]{MaDiJa06}.

Neutrino transport is computed in the ``ray-by-ray-plus'' (RbR+) approximation \cite{BuRaJa03}, where an independent, spherically symmetric transport solve is computed for each ``ray'' (radial array of zones with the same $\theta$, $\phi$). (It is very important to note that the RbR+ approximation does not restrict the neutrinos to strict radial propagation only. In spherical symmetry, neutrinos propagate along arbitrary rays, not just radial rays, but the {\em net} angular flux is zero, leaving only radial flux. Each RbR+ solve is a {\em full} spherically symmetric solve (see Figure \ref{fig:rbr}). The 3D problem is broken up into $N_{\theta}\times N_{\phi}$ spherically symmetric problems, where $N_{\theta,\phi}$ are the number of latitudinal and longitudinal zones, respectively. RbR+ is exact (physically speaking, modulo numerical error) if the neutrino source is spherically symmetric. Thus, if accreted material raining down on the PNS surface via the non-spherical accretion funnels, obvious in Figures~\ref{fig:entropy} and \ref{fig:entropy3D}, and creating hot spots, spreads rapidly over the surface relative to the neutrino-heating and shock-revival time scales, which we find it does, and in the absence of significant rotation, the RbR+ approximation is a reasonable approximation, at least initially. There are practical benefits to the approximation, as well, which we will discuss later.)

The transport solver for each ray is an improved and updated version of the multi-group flux-limited diffusion transport solver of Bruenn \cite{Brue85} enhanced for GR \cite{BrDeMe01}, with an additional geometric flux limiter to prevent an overly-rapid transition to free streaming of the standard flux-limiter.  All $O(v/c)$ observer correction terms have been included.

\chimera\ solves for all three flavors of neutrinos and antineutrinos with four coupled species: \nue, \nuebar, $\numt=\{\numu,\nutau\}$, $\numtbar=\{\numubar,\nutaubar\}$, with typically 20 energy groups covering two decades in neutrino energy.
Our standard, modernized, neutrino--matter interactions include emission, absorption, and non-isoenergetic scattering on free nucleons \cite{RePrLa98}, with weak magnetism corrections \cite{Horo02}; emission/absorption (electron capture) on nuclei \cite{LaMaSa03}; isoenergetic scattering on nuclei, including ion-ion correlations; non-isoenergetic scattering on electrons and positrons; and pair emission from $e^+e^-$-annihilation \cite{Brue85} and nucleon-nucleon bremsstrahlung \cite{HaRa98}.
\chimera\ generally utilizes the $K = 220$~\mev\ incompressibility version of the Lattimer--Swesty \cite{LaSw91} EoS for  $\rho>10^{11}\,\gcc$ and a modified version of the Cooperstein \cite{Coop85} EoS for  $\rho<10^{11}\,\gcc$, where nuclear statistical equilibrium (NSE) applies.
Most \chimera\ simulations have used a 14-species $\alpha$-network ($\alpha$, \isotope{C}{12}-\isotope{Zn}{60}) for the non-NSE regions \cite{HiTh99a}. In addition,
\chimera\ utilizes a 17-nuclear-species NSE calculation for the nuclear component of the EOS for $Y_{\rm e}>26/56$ to provide a smooth join with the non-NSE regime

During evolution, the radial zones are gradually and automatically repositioned to track changes in the mean radial structure.
To minimize restrictions on the time step from the Courant limit, the lateral hydrodynamics for a few inner zones are ``frozen'' during collapse, and after prompt convection fades, the laterally frozen region expands to the inner 6--8~km.
In the ``frozen'' region the  radial hydrodynamics and neutrino transport are  computed in spherical symmetry.

The supernova code most closely resembling \chimera\ 
is the \vertex\ code developed by the Max Planck group \cite{BuRaJa03,BuRaJa06,BuJaRa06,MuJaDi10}. This code utilizes a RbR+ approach to neutrino transport, solving the first two multifrequency angular moments of the transport equations with a variable Eddington closure that is solved at intervals using a 1D approximate Boltzmann equation.

\chimera\ does not yet include magnetic fields. Studies with \chimera\ that include magnetic fields will be part of future efforts. 

\section{Our Approach in Context}

\begin{figure}
\includegraphics[width=3.25in]{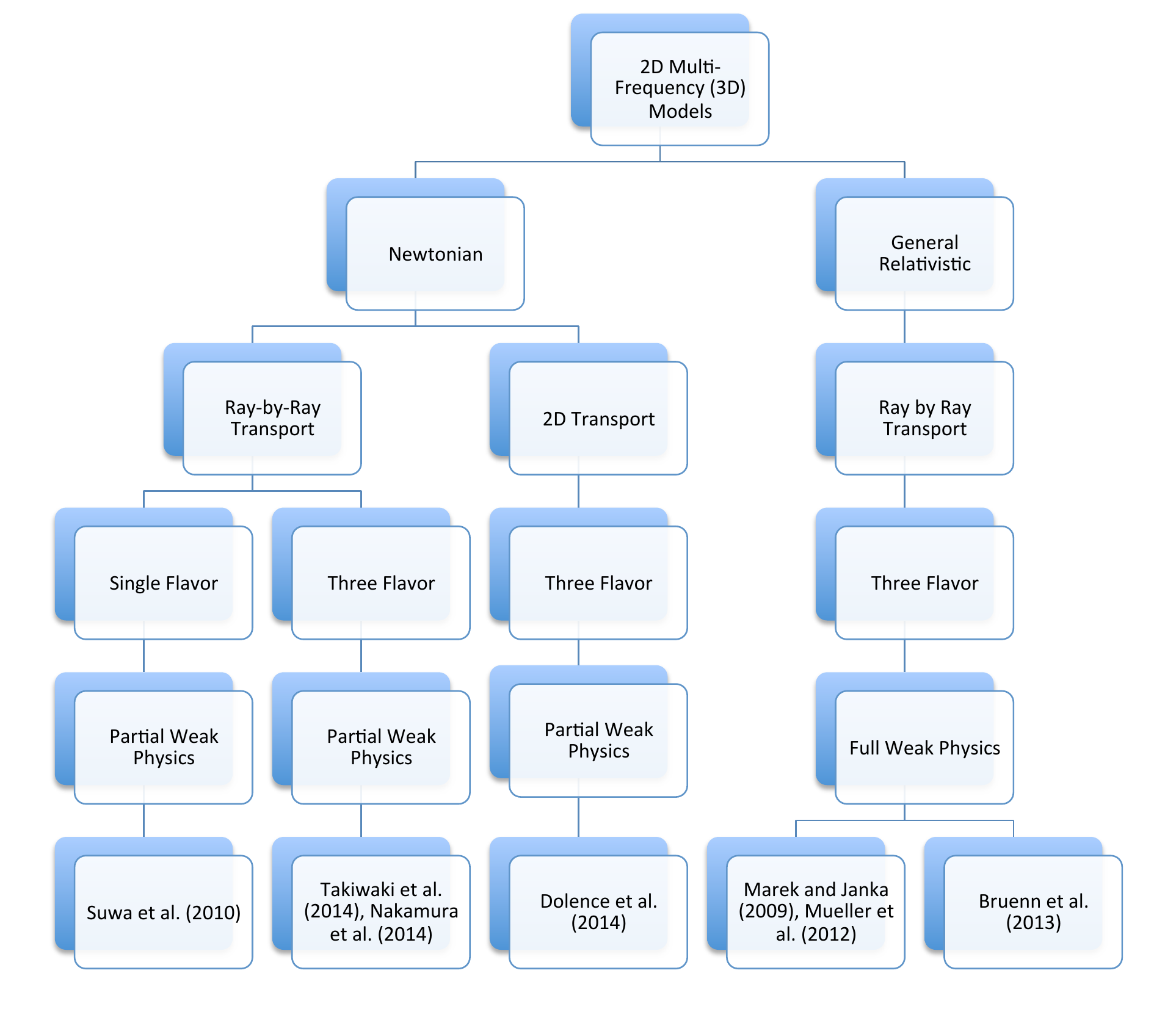}
\caption{An overview of the approaches used in the context of 2D CCSN modeling by various groups around the world \cite{SuKoTa10,TaKoSu14,NaTaKu14,DoBuZh14,maja09,MuJaMa12,BrMeHi13}.  
\label{fig:label2DApproaches}}
\end{figure}

\begin{figure}
\includegraphics[width=3.0in]{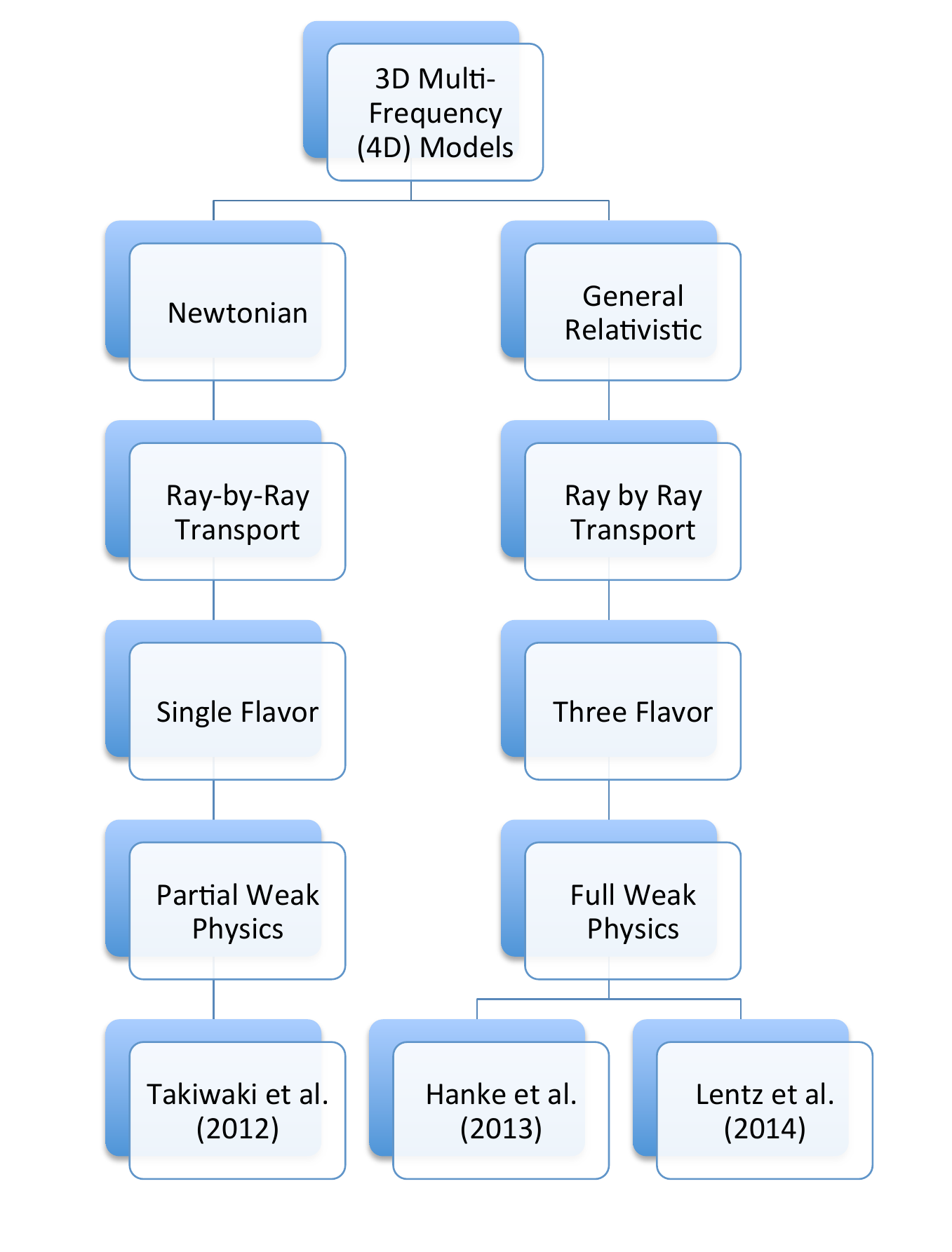}
\caption{An overview of the approaches used in the context of 3D CCSN modeling by several groups around the world \cite{TaKoSu12,HaMuWo13,LeBrHi15}.
\label{fig:label3DApproaches}}
\end{figure}

A number of 2D simulations have been performed to date with multi-frequency neutrino transport. These break down into two classes, those that have implemented the RbR neutrino transport approximation and those that have not --- i.e., those that have implemented 2D transport. Figure \ref{fig:label2DApproaches} provides an overview of the approaches used by various supernova groups in producing these 2D models. It is clear the RbR approximation has enabled the inclusion of general relativity and state-of-the-art neutrino interactions, at the expense of the added spatial dimensionality of the transport, whereas the non-RbR approach includes the second spatial dimension in the neutrino transport, but does so at the expense of realism in the treatment of gravity and the neutrino interactions with stellar matter. The reason for this is simple: In the RbR approach, transport codes that have been used in spherically symmetric studies, such as \AgileBOLTZTRAN\ , can be deployed. These codes already, or at least can more easily, include all relativistic transport corrections and full weak interaction physics. To achieve the same level of sophistication in two and three spatial dimensions is more difficult and far more computationally intensive. For example, a 3D multi-frequency approach (e.g., flux-limited diffusion or a variable Eddington tensor method) will require the sustained-petaflop performance of present-day leadership-class computing facilities. In light of the practical difficulties associated with including more physics in fully 3D simulations, the RbR approximation provides an alternative approach that can be used in the interim. The use of both approaches by the community as it moves forward will be essential, as simulations with RbR neutrino transport with approximate general relativity and full weak interaction physics must be gauged by non-RbR approaches that can test the efficacy of the RbR approach. Ultimately, the two approaches must merge, with 3D simulations performed with 3D (i.e., not RbR) general relativistic neutrino transport, general relativistic hydrodynamics and gravity, and a full weak interaction set. Figure \ref{fig:label3DApproaches} gives an overview of the 3D simulations performed to date, using multi-frequency neutrino transport. It is obvious that fewer groups have attempted this, and far fewer simulations have been performed. It is also evident they have all been performed with RbR and not 3D neutrino transport.

\section{Results from our 2D Core Collapse Supernova Models}\label{sec:current2D}

\begin{figure}
\includegraphics[width=3.5in]{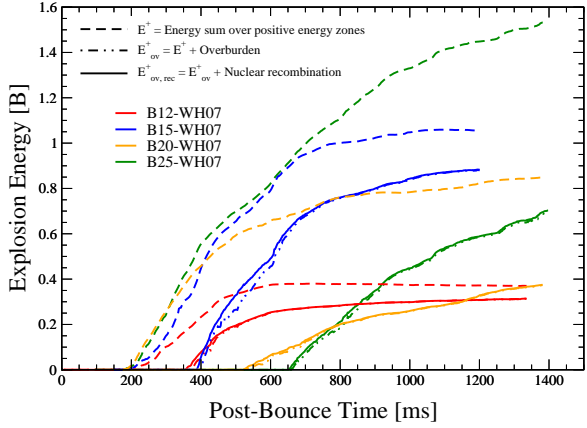}
\caption{Diagnostic energy (\Ediag; dashed lines) versus post-bounce time for all of our published 2D models \cite{BrMeHi13,BrLeHi14}. Dash-dotted lines (\Ediagov) include binding energy of overburden and dashed lines (\Ediagovrec) also include estimated energy gain from nuclear recombination.}
\label{fig:energy}
\end{figure}

We \cite{BrMeHi13,BrLeHi14} have performed four 2D simulations with \chimera\ beginning with the 12, 15, 20, and 
25~\msun\ progenitors of Woosley and Heger \cite{wohe07}.
One result of these simulations is the realization that a fully developed (and therefore final) explosion energy will require much more lengthy simulations than anticipated in the past.
In the explosion energy plot, Figure~\ref{fig:energy}, the dashed lines show the growth of the ``diagnostic energy'' (the sum of the gravitational potential energy, the kinetic energy, and the internal energy in each zone --- i.e., the total energy in each zone --- for all zones having a total energy greater than zero) along with more refined estimates of the final explosion energy that account for the work required to lift the as-yet-unshocked envelope ``overburden'' (dash-dotted lines) and, in addition, the estimated energy released from recombination of free nucleons and alpha particles into heavier nuclei (solid lines). We expect these latter two measures to bracket the final kinetic energy of the fully developed explosion. Using the definition of the explosion energy that includes both the energy cost to lift the overlying material and the energy gain associated with nuclear recombination, we can define $t_{\rm explosion}$, the explosion time, which is the time at which the explosion energy becomes positive and, therefore, the explosion can be said to have been initiated. For the 12, 15, 20, and 25 M$_\odot$ models, $t_{\rm explosion}$ is approximately 320, 320, 500, and 620 ms after bounce, respectively. 

\begin{figure}
\includegraphics[width=3.00in]{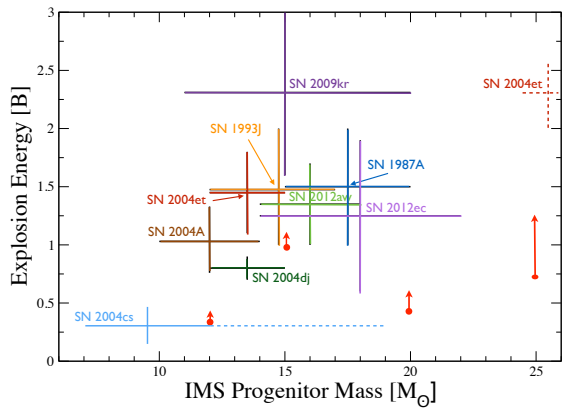}
\caption{
Observed explosion energies for a number of CCSNe, along with predicted explosion energies from our 12, 15, 20, and 25 M$_\odot$ progenitor models (red dots) \cite{BrLeHi14}. The arrows indicate that our explosion energies are still increasing at the end of each run. The length of each arrow is a measure of the rate of change of the explosion energy at the end of the corresponding run.
\label{fig:energycomparison}
}
\end{figure}

\begin{figure}
\includegraphics[width=3.00in]{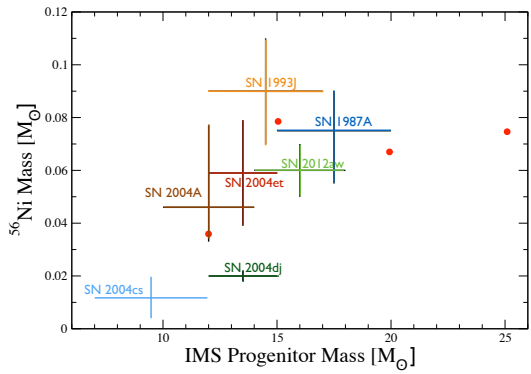}
\caption{
Observed production of $^{56}$Ni for a number of CCSNe, along with our predictions from our 12, 15, 20, and 25 M$_\odot$ progenitor models (red dots) \cite{BrLeHi14}.
\label{fig:nickelmass}
}
\end{figure}

\begin{figure}
\includegraphics[width=3.00in]{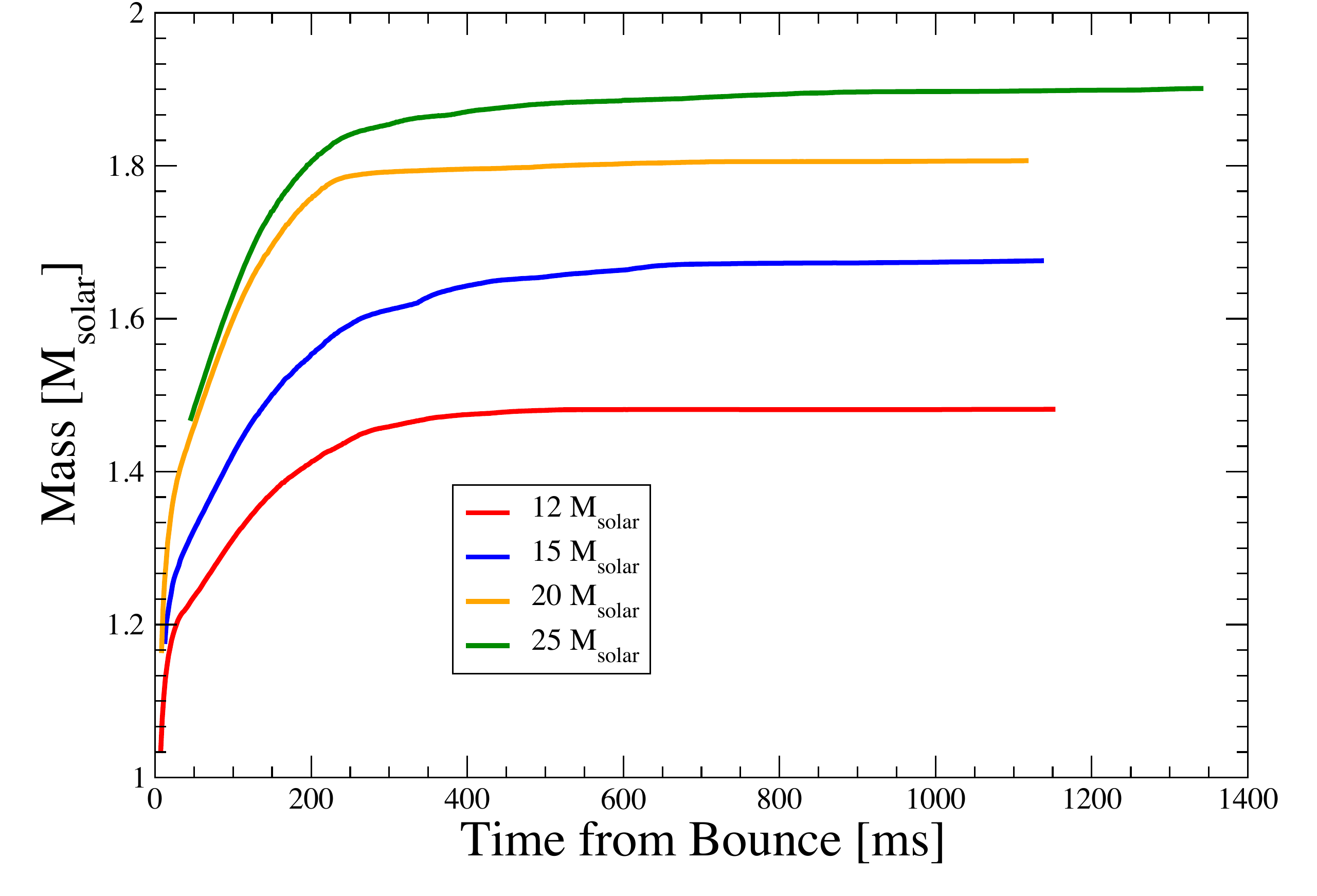}
\caption{
Time evolution of the proto-neutron star (baryonic) mass in each of our 4 2D models, beginning with 12, 15, 20, and 25 M$_\odot$ progenitors \cite{BrLeHi14}.
\label{fig:pnsmass}
}
\end{figure}

Moving now to a comparison with observations: All four models have achieved explosion energies that are in the $\approx $0.4--1.4 Bethe range of observed Type~II supernovae (see Figure \ref{fig:energycomparison}). Figures \ref{fig:nickelmass} and \ref{fig:pnsmass} compare our predictions for the mass of $^{56}$Ni produced and the final proto-neutron star (baryonic) masses produced, respectively, with observations. Note, the large systematic errors in observed progenitor masses preclude any detailed comparison between our results and observations {\em as a function of progenitor mass}. Nonetheless, comparisons of our predicted {\em ranges} of explosion energies, $^{56}$Ni masses, etc. with observed ranges is meaningful and demonstrates we are making progress toward developing predictive models.

\begin{figure}
\includegraphics[width=3.25in]{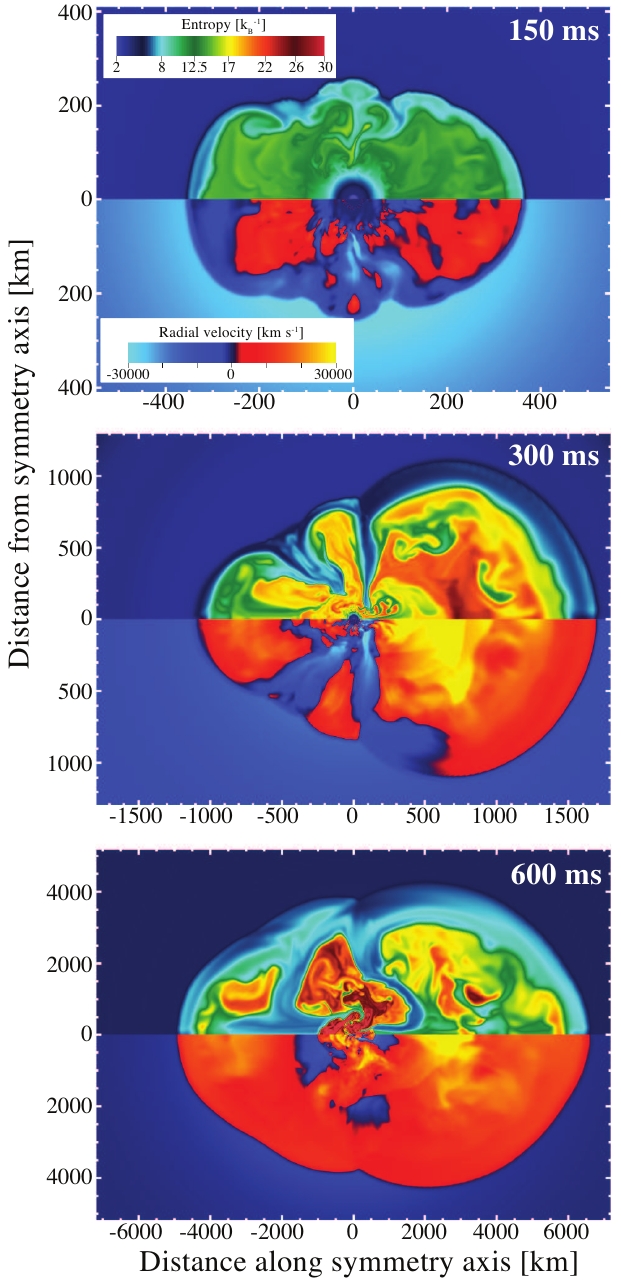}
\caption{Evolution of the entropy (upper half) and radial velocity (lower half) at 150, 300, and 600~ms after bounce  for the 12~\msun\ model of Bruenn et al. \cite{BrMeHi13}.  
\label{fig:entropy}}
\end{figure}

Three snapshots of hydrodynamic motion are visible in 
Figure~\ref{fig:entropy}, 
which shows the entropy (upper half) and radial velocity (lower half) for the 12 \msun\ model at 150~ms, 300~ms, and 600~ms after bounce.  
At 150~ms, roughly 100~ms before rapid shock expansion heralds the onset of a developing explosion, asphericity is developing as a result of vigorous neutrino-driven convection and the SASI.  
By 300~ms large-scale, high-entropy, buoyant plumes are evident, as the explosion continues to develop.  
However, low-entropy down-flows still connect the unshocked regions with the PNS surface, continuing to supply accretion energy to power the neutrino luminosities driving the development of the explosion. By 600~ms, these down-flows have been cut off by the expanding ejecta, but their remnants continue to accrete onto the PNS, allowing the explosion to continue to gain in strength.

Though these simulations have run further into explosion than previous simulations, the final explosion energies --- in particular, for the 20 and 25 M$_\odot$ models --- are clearly still developing. 
These simulations will therefore continue. Additional 2D simulations --- e.g., using different progenitor masses --- are planned.

\section{Preliminary Results from our 3D Core Collapse Supernova Model}\label{sec:current3D}

\begin{figure}
\includegraphics[width=3.1in]{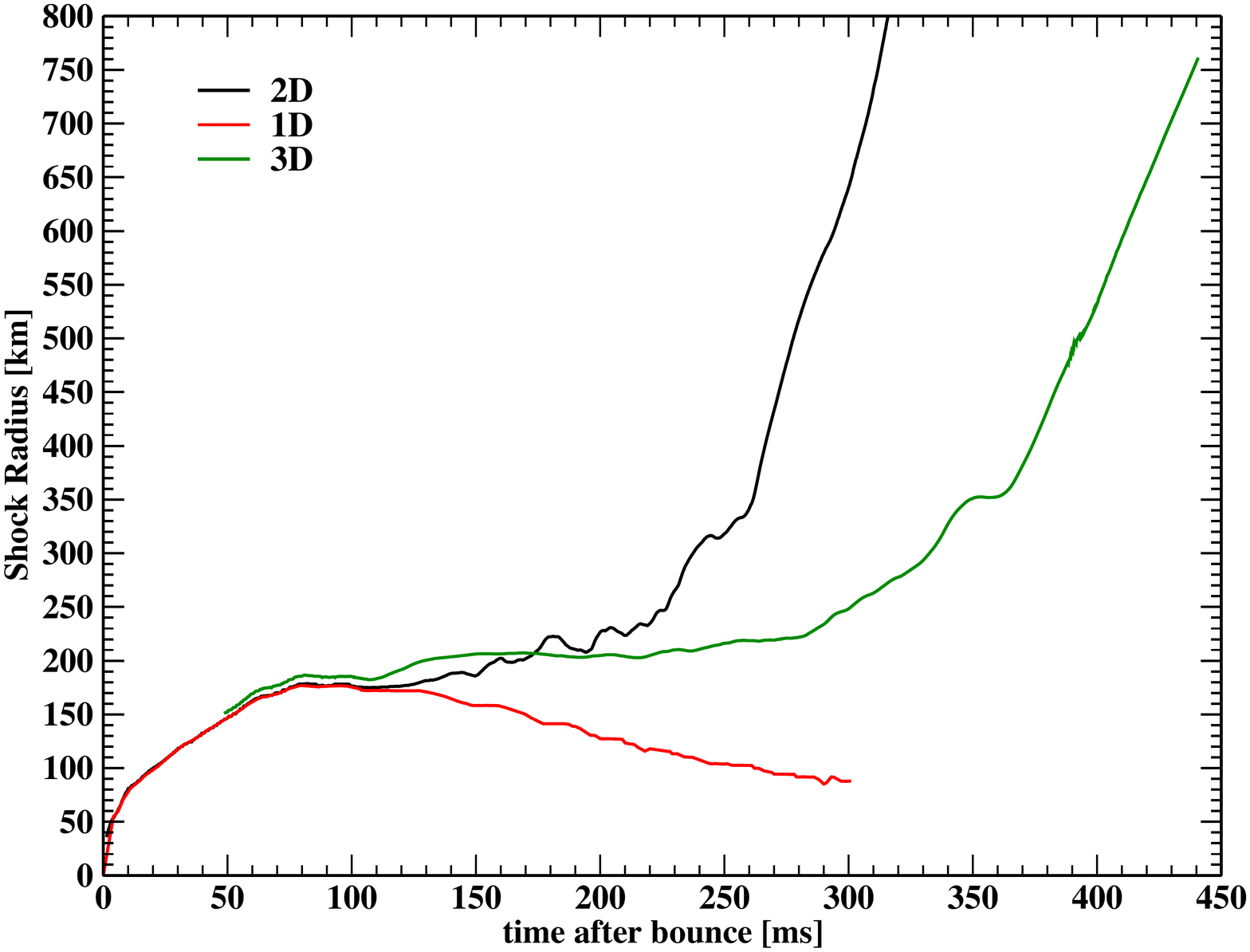}
\caption{Evolution of the shock trajectory from our 1D model and the angle-averaged shock trajectories from our 2D and 3D models, all for the 15~\msun\ case \cite{LeBrHi15}. The 1D model does not develop an explosion, whereas an explosion is obtained in both our 2D and our 3D models.
\label{fig:1D2D3DShockTrajectories}
}
\end{figure}

\begin{figure}
\includegraphics[width=3.15in]{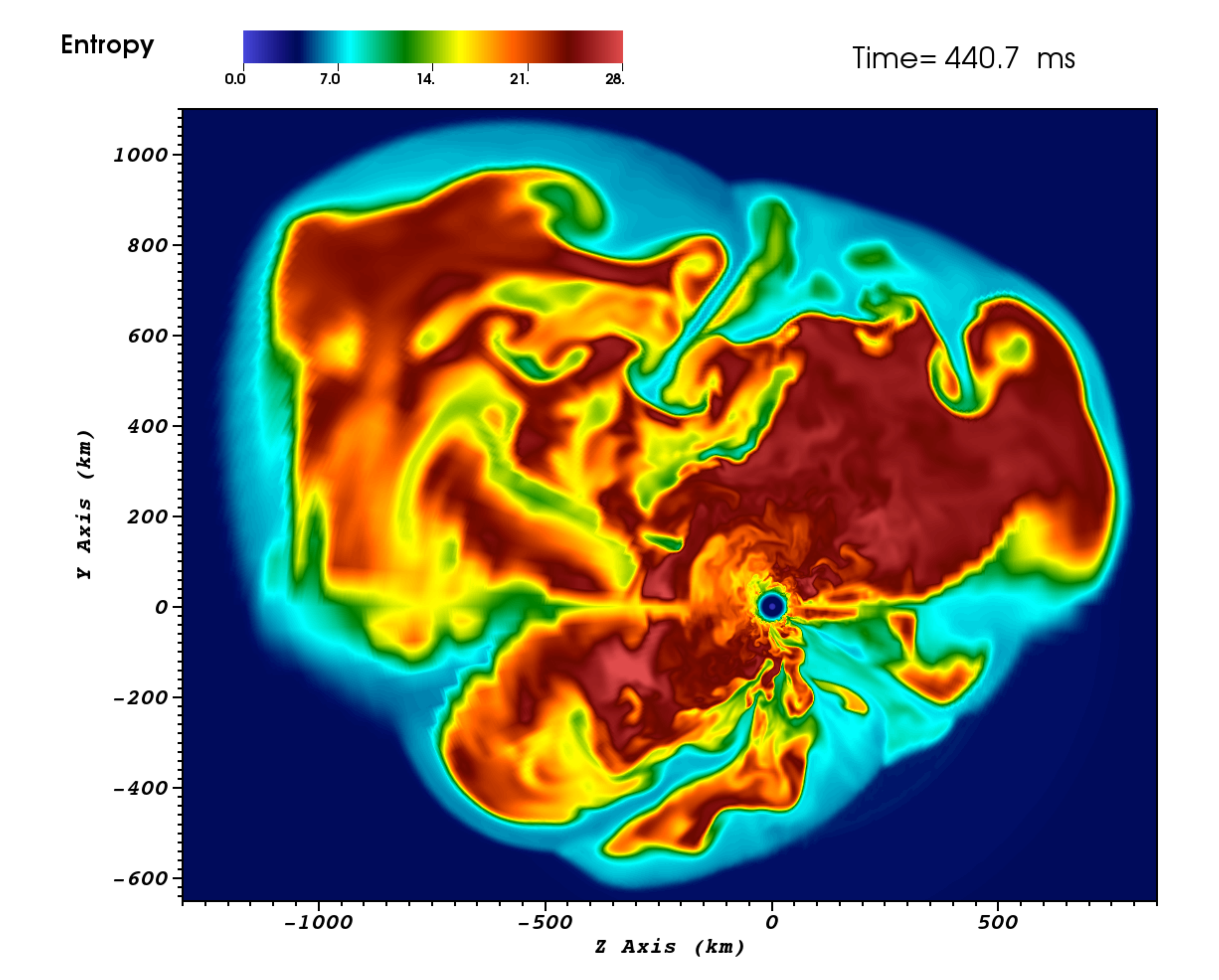}
\caption{Snapshot of the equatorial cross section of the entropy in our ongoing 3D simulation for the 15~\msun\ case at $\sim$441 ms after bounce \cite{LeBrHi15}. Red indicates high-entropy, expanding, rising material. Green/blue indicates cooler, denser material. Evident are significant (green) down flows fueling the neutrino luminosities.
\label{fig:entropy3D}
}
\end{figure}

Few 3D multiphysics models with necessary realism (as defined above) have been performed. Notable among these is the recently published model of Hanke et al. \cite{HaMuWo13}. Preliminary results from the Oak Ridge group \cite{LeBrHi15} in the context of a model similar to the Garching group's model -- i.e., with essentially the same physics and treatment of this physics -- are presented here, although we begin with the same 15 M$_\odot$ Woosley--Heger progenitor used in our 2D models, whereas they began with the 27 M$_\odot$ Woosley--Heger progenitor. 

Figure \ref{fig:1D2D3DShockTrajectories} shows the angle-averaged shock trajectories from our one-, two-, and three-dimensional models, all run with the \chimera\ code beginning with the same 15 M$_\odot$ Woosley--Heger progenitor and including the same (full) physics. Explosion is evident in both the 2D and the 3D cases. Explosion is not obtained in 1D. Comparing the two- and three-dimensional trajectories, we see that the development of the explosion in the 3D case is slower. In the 2D case, the shock radius changes rapidly beginning at about 200 ms after bounce. In the 3D case, the shock radius does not begin to climb dramatically until approximately 100 ms later, at $\sim$300 ms after bounce. The 1D and 2D/3D angle-averaged shock radii diverge at approximately 125 ms after bounce, and the 2D and 3D angle-averaged shock radii diverge later, at about 200 ms after bounce.

Figure \ref{fig:entropy3D} is a snapshot of a 2D slice of our ongoing 3D model at approximately 441 ms after bounce. Shown is the stellar core entropy. The shock wave is clearly outlined by the jump in entropy across it. Neutrino-driven convection is evident in the slice. Hotter (red) rising plumes bring neutrino-heated material up to the shock, while cooler (green) down flows replace the fluid below. Distortion of the shock away from axisymmetry and the nonaxisymmetric patterns of convection beneath the shock are also evident. Conclusive evidence for $l=1$, ``sloshing'' and $m=1$, ``spiral'' modes of the SASI will require a modal analysis, although the 2D slice clearly does not rule out either mode. 

This simulation utilizes 32,400 rays (solid angle elements) with 2\degree\ resolution in longitude and a resolution in latitude that varies from 8\degree\ at the pole to better than 0.7\degree\ at the equator, but is uniform in the cosine of the colatitude. 
Due to the Courant limit, the coordinate pole in standard spherical-polar coordinates creates a strong restriction on the time step size and therefore lengthens the total run time compared to a similar resolution 2D simulation.  
Our constant cosine-of-colatitude grid seeks to minimize this impact without resorting to a grid that is coarse at all latitudes or implementing unevolved (frozen) regions near the pole. The simulation will consume approximately 100 M core--hours to complete. {\em (This gives a strong indication of how the physics included in the models, even in the RbR+ approximation, significantly drives upward their computational cost.)}
As this 3D simulation for a 15~\msun\ progenitor evolves, we will be able to examine the nature of the CCSN explosion mechanism without the assumption of axisymmetry that is inherent in the 2D models. {\em The} key question: Will this model yield a robust explosion? And will other predictions agree with observations? As indicated by all of our 2D models, our current 3D model will need to be run significantly longer, and detailed computations of the explosion energy and other observables will need to be completed before we can begin to answer these questions.

\section{Conclusions and Outlook}

The most sophisticated spherically symmetric models developed to date do not exhibit core collapse supernova explosions. Despite the prodigious amount of gravitational binding energy tapped during stellar core collapse and radiated via neutrinos, neutrino heating of the stellar core material beneath the supernova shock wave, unaided by other physics, is unable to power such explosions. On the other hand, with the aid of neutrino-driven convection beneath the shock, and the SASI, robust explosions have been obtained in both two- and three-dimensional models, with model predictions consistent with observations of multiple quantities (explosion energy, $^{56}$Ni mass, neutron star mass, neutron star kick velocity).

One- and two-dimensional studies have identified a list of key physics needed in CCSN models. The addition of new physics (e.g., magnetic fields) will likely add to this list as the new physics is added to today's most advanced models (e.g., see \cite{ObJaAl14}). It is also possible that the addition of new physics will render some of the physics currently included less important. However, it is unlikely that the impact of general relativity and of important neutrino physics (e.g., relativistic transport corrections such as gravitational redshift and the full physics of electron capture and neutrino scattering) will be significantly lessened by adding new physics. The quantum leap in CCSN modeling that occurred two decades ago, where axisymmetry replaced spherical symmetry, did not reduce the importance of this physics --- case in point, both Lentz et al. \cite{LeMeMe12} and Mueller et al. \cite{MuJaMa12} reached the same conclusions. Moreover, the development of magnetic fields will depend on the environment established by accretion and neutrino heating.
Future modeling --- in particular, the direction we choose to take --- should rely on the predictions of the best {\em available} models, more so than on speculation of what physics may or may not be important. With this in mind, the task at hand is, therefore, to build 3D models with the minimum physics set identified in the studies mentioned above. 

In this brief review, we outlined the approaches used by the various supernova modeling groups around the world, focusing on two- and three-dimensional, multi-frequency models. While a comparative analysis of the results of these studies can shed light on the impact of (a) Newtonian versus general relativistic gravity, hydrodynamics, and neutrino transport, and/or (b) including a reduced versus a complete set of neutrino weak interactions, the latter of which would include detailed nuclear electron capture and neutrino energy scattering, results from simulations cutting across these various levels of sophistication should not be compared with the expectation that the outcomes --- in particular, whether or not robust explosions are obtained --- should be the same. For example, comparing a Newtonian and a general relativistic model, with all other physics in the models kept the same, allows us to understand the role of general relativity, but we should not expect the Newtonian and general relativistic models to agree quantitatively, or even qualitatively.

Having said this, a comparison between, for example, the results obtained by the Oak Ridge and Garching groups can be made given the similarity of their approaches and the physics included in each of their model sets. In this context, it is important to note that the results of the Garching group differ between simulations performed with their \vertex\ code \cite{maja09}, which uses a general relativistic monopole correction to the Newtonian self-gravitational potential, derived from the Tolman-Oppenheimer-Volkov equation of the spherically-averaged fluid and thermodynamic quantities in the stellar core, and with their \coconutvertex\ code \cite{MuJaMa12}, which instead uses the conformal flatness approximation to the general relativistic gravitational field. \vertex\ is the code most similar to \chimera\ . Unfortunately, to date, results from the \vertex\ code using the more modern Woosley--Heger progenitor set \cite{wohe07} have not been published, so a direct comparison is not yet possible.

Focusing once again on the ongoing 3D simulations cited here: Will robust neutrino-driven explosions be obtained? If the answer is no, three explanations are possible: (1) Removing current approximations in the models (e.g., the use of RbR neutrino transport) and/or making other improvements (e.g., increasing the spatial resolution) may fundamentally alter the outcomes. (2) We are missing essential physics. (3) A combination of additional physics and improved modeling may be needed to alter the outcomes. 
With regard to (1)-(3):

(A) All of the simulations documented here were initiated from state-of-the-art (e.g., the \citet{wohe07} series) spherically-symmetric progenitor models. 
Couch and Ott \cite{CoOt13} point out that multidimensional simulations of the advanced stages of stellar evolution of massive stars yield large deviations from 
spherical symmetry in the Si/O layer (see \cite{Arnett14} and the references cited therein).
They demonstrate that such (expected) deviations from spherical symmetry can qualitatively alter the 
post-stellar-core-bounce evolution, triggering an explosion in a model that otherwise fails to explode. Such a qualitative change in outcome 
demands better initial conditions, which can be obtained when spherically symmetric models, currently able to complete stellar evolution through 
silicon burning and the formation of the iron core (multidimensional models are not yet capable of this), are informed by 3D stellar
evolution models of earlier burning stages.

(B) Given the importance of the SASI in the explosion models developed thus far, and given that the SASI is a long-wavelength instability, how will the SASI and the turbulence it induces, or neutrino-driven convection and the turbulence it induces, interact? There is evidence, for example, that the energy in long-wavelength modes of the SASI are sapped by the very turbulence the SASI seeds, as a result of the significant shear between counterrotating flows induced by its $m=1$ spiral mode in three dimensions \citep{EnCaBu12}. On the other hand, Couch and Ott \cite{CoOt14} recently showed that turbulent ram pressure may be important in driving the shock outward, relieving some of the work from the thermal pressure associated with neutrino heating. Moreover, significant deviations from spherical symmetry in the progenitor, as would be expected based on the current 3D stellar evolution models discussed above, would seed turbulence and, thus, potentially enhance the contribution of turbulence to the outward pressure driving the shock.

(C) If we maintain that CCSNe are neutrino-driven, it may be logical to assume that we are missing something essential in the neutrino sector. Motivated by the experimental and observational measurement of neutrino mass, recent efforts to explore its impact on neutrino transport in stellar cores have uncovered new and increasingly complex physical scenarios \citep{dufuqi10,chcafr12,chcafr13,VlFuCi14}. Now that the quantum kinetic equations for neutrinos in stellar cores have been derived (e.g., see \cite{VlFuCi14}), efforts can begin in earnest to extend Boltzmann models to include the quantum mechanical coherent effects associated with neutrino mass. This is, of course, a long-term goal. It is not clear that physics associated with neutrino mass will have an impact on the explosion mechanism, but it has been demonstrated that such physics may impact terrestrial CCSN neutrino signatures significantly (e.g., see \cite{DuFuCa07}).

Since Colgate and White first proposed that CCSNe are neutrino-driven \cite{CoWh66}, nearly five decades have passed. Ascertaining the CCSN explosion mechanism has certainly been a challenge. Each new piece of physics, each new dimension, has brought both breakthroughs and additional challenges. Nonetheless, the last decade of CCSN modeling has led to rapid progress. This progress --- in particular, the recent progress outlined here --- and the growing capability of available supercomputing platforms, encourage us that a solution to this long-standing astrophysics problem is achievable with a continued, systematic effort in perhaps the not-too-distant future.


\end{document}